# The Rest Mass of the Hydrogen Atom from First Principles

Ernst Karl Kunst

**The rest mass of the hydrogen ($^1$H) atom in its ground state is calculated from first physical principles and elementary geometric considerations.**

Previously has been shown [1] that rest mass "m" and relativistic mass m' = m$\gamma_0$ - where $\gamma_0$ is the Lorentz factor $(1 - v_0^2/c^2)^{-1/2}$ based on composite velocity $v_0$ [2] - must be of like origin (are equivalent) and hence the former seems to be generated by the movement of a fourth spatial dimension of matter relative to a fourth dimensional manifold $R_4$, in which our $R_3$-world is embedded. This implies that - apart from a numerical factor - the following must be valid

$$m_H = \frac{\lambda_4}{c} = \frac{2\lambda_1}{c} = 2\tau_1 = \frac{\sqrt{h}}{c} \, ,$$

where $m_H$ presumably is the rest mass of the hydrogen atom in its ground state, $\lambda_4$ fundamental length in $R_4$ and $\lambda_1$ in $R_1$, respectively, $\tau_1 = \lambda_1/c$ quantum of time, h Planck's constant and c velocity of light.

It is proposed to calculate the mentioned numerical factor as follows.

By development of energy E' = E$\gamma_0$ = E$(1 - v_0^2/c^2)^{-1/2}$ to powers of $v_0^2/c^2$ one receives

$$E' = mc^2 \left( 1 + \frac{v_0^2}{2c^2} + \frac{3v_0^4}{8c^4} \right) ,$$

where E means rest energy. As is widely known does the first term mc$^2$ of the right-hand side express the rest mass and the second term m$v_0^2$/2 the classical kinetic energy of the material particle under consideration. Thus, the third term $3mv_0^4/(8c^2)$ must be the energy, which is due alone to the relativistic expansion of the moving material particle [2]. Therefore, if rest mass is generated by the movement of $\lambda_4$ relative to $R_4$ at velocity c it must be valid $3mc^2/8 = \lambda_4/c$ or

$$\frac{3 V_m \rho_m c^2}{8} = \lambda_4 c, \qquad\qquad (1)$$

where $V_m$ is volume and $\rho_m$ Newtonian density of mass. It is to expect that the volume $V_H$ of the hydrogen atom in $R_3$ attains the minimum value of volume unit 1, which is the volume of the tetraoid formed by the four points 1; 2; 3; 4 with the coordinates $x_1$, $y_1$, $z_1$;...;$x_4$, $y_4$, $z_4$:

$$(1, 2, 3, 4) = \frac{1}{1 \times 2 \times 3} \begin{vmatrix} x_1 & y_1 & z_1 & 1 \\ x_2 & y_2 & z_2 & 1 \\ x_3 & y_3 & z_3 & 1 \\ x_4 & y_4 & z_4 & 1 \end{vmatrix}$$

Thus, if furthermore $V_H = dx_1 dx_2 dx_3$, from (1) follows

$$6 \times \frac{\sqrt[3]{3} \, dx_1}{2} \times \frac{\sqrt[3]{3} \, dx_2}{2} \times \frac{\sqrt[3]{3} \, dx_3}{2} \times \rho_m = \frac{\lambda_4}{c} . \qquad (2)$$

From the foregoing it is clear that $m = V_m \rho_m$ is a four dimensional object so that must be valid $\rho_m = \sqrt[3]{3} dx_4/(2c)$. Thus (2) delivers

$$6 \times \frac{3}{8} \times \frac{\sqrt[3]{3}}{2} \times V_H \frac{dx_4}{c} = \frac{\lambda_4}{c}$$

or

$$V_H \frac{dx_4}{c} = V_H \rho_H = m_H = \frac{8 \lambda_4}{9 \sqrt[3]{3} \, c} = \frac{8 \sqrt{h}}{9 \sqrt[3]{3} \, c} . \qquad (3)$$

Equ. (3) delivers for the rest mass of the [1]H-atom

$$m_H = 1.673456 \times 10^{-24} \ g , \qquad (4)$$

which value agrees nearly, but not exactly with the experimental value $1{,}673559 \times 10^{-24}$ g [3] - g means gram. As has been shown before [4] does the gravitational field of a body contribute to its inertial as well as to ist gravitational mass, namely

$$M = m + m_V = m \left( 1 + \frac{2 G(\sqrt{R} - \sqrt{R_1})}{\sqrt[3]{c}} \right) , \qquad (5)$$

where M is the sum of the mass (of a body) and of the surrounding gravitational field;, $m_V$ the mass of the gravitational field, G the gravitatonal constant and R the radial distance from the center of the body. Therefore, the integrated mass of the sub-microscopic [1]H-atom in our medioscopic world must be composite of the central rest mass (3) of the atom and the surrounding (gravitational) field vacuum. The mass of the gravitational field cannot be calculated straightforwardly, because there is no means to determine a boundary of R. But if the mass of the field vacuum of the ¹H-atom is compared with the mass of the gravitational field of a body of equal density but different mass, according to (5) the ratios

$$\frac{m_{V_H}}{m_V} = \frac{m_H \sqrt{R_{V_H}}}{m \sqrt{R_V}} ,$$

and

$$\frac{R_{V_H}}{R_V} = \frac{R_H}{R} ;$$

result, where m means mass of the body of comparison and $m_V$ the mass of its gravitational field. If m = $R_V$ = 1, implying $m_V$ = 1, we receive

$$m_{V_H} = m_H \sqrt{R_H} , \tag{6}$$

where $R_H$ is the radius of the atom. Hence the global mass of the ¹H-atom, including the mass of its gravitational field, according to (3) to (6) is given by

$$M_H = m_H \left( 1 + \sqrt{R_H} \sqrt[6]{\frac{3}{8}} \right) = \frac{8\sqrt{h}}{9\sqrt[3]{3}\,c} \left( 1 + \sqrt{a_0} \sqrt[6]{\frac{3}{8}} \right) , \tag{7}$$

where $R_H = a_0$ is the first Bohrian radius. Calculation delivers 1,673559 × 10⁻²⁴ g, which fits exactly the experimental result [3].
It seems that (7) denotes the exact value of the rest mass of the smallest possible, electrically neutral and durable piece of matter - as seen from our medioscopic level. Of course, the theory delivers no explanation yet of the masses of the elementary particles. Presumably those masses are due to a hitherto not yet understood layered structure of space-time on the sub-microscopic level.

Dedicated to Barbara.